# One year of COVID-19 vaccine misinformation on Twitter: Longitudinal study

Francesco Pierri[1,2,*], Matthew R. DeVerna[2], Kai-Cheng Yang[2], David Axelrod[2], John Bryden[2], Filippo Menczer[2]

**1** Dipartimento di Elettronica, Informazione e Bioingegneria, Politecnico di Milano, Italia
**2** Observatory on Social Media, Indiana University, Bloomington, USA

*francesco.pierri@polimi.it

## Background

Vaccinations play a critical role in mitigating the impact of COVID-19 and other diseases. Past research has linked misinformation to increased hesitancy and lower vaccination rates. Gaps remain in our knowledge about the main drivers of vaccine misinformation on social media and effective ways to intervene.

## Objective

Our longitudinal study has two primary objectives: (i) to investigate the patterns of prevalence and contagion of COVID-19 vaccine misinformation on Twitter in 2021; and (ii) to identify the main spreaders of vaccine misinformation. Given our results, we also question what are the likely drivers of misinformation and its spread, providing insights for potential interventions.

## Methods

We collected almost 300M English-language tweets related to COVID-19 vaccines using a list of over 80 relevant keywords over a period of 12 months. We then extracted and labeled news articles at the source level, based on third-party lists of low-credibility and mainstream news sources, and measured the prevalence of different kinds of information. We also considered suspicious YouTube videos shared on Twitter. We focused our analysis of vaccine misinformation spreaders on verified and automated Twitter accounts.

## Results

Our findings show a relatively low prevalence of low-credibility information compared to the entirety of mainstream news. However, the most popular low-credibility sources had reshare volumes comparable to many mainstream sources, and larger volumes than authoritative sources such as the U.S. Centers for Disease Control and Prevention and the World Health Organization. Throughout the year, we observed an increasing trend in the prevalence of low-credibility news about vaccines. We also observed a considerable amount of suspicious YouTube videos shared on Twitter. Tweets by a small group of about 800 "superspreaders" verified by Twitter accounted for approximately 35% of all reshares of misinformation on an average day, with the top superspreader (*@RobertKennedyJr*) responsible for over 13% of retweets. Finally, low-credibility news and suspicious YouTube videos were more likely to be shared by automated accounts.

## Conclusions

The wide spread of misinformation around COVID-19 vaccines on Twitter during 2021 shows that there was an audience for this type of content. Our findings are also consistent with the hypothesis that superspreaders are driven by financial incentives that allow them to profit from health misinformation. Despite high-profile cases of deplatformed misinformation superspreaders, our results show that in 2021 a few individuals still played an outsize role in the spread of low-credibility vaccine content. As a result, social media moderation efforts would be better served by focusing on reducing the online visibility of repeat-spreaders of harmful content, especially during public health crises.

# Introduction

The global spread of the novel coronavirus (SARS-CoV-2) over the last two years affected the lives of most people around the world. As of December 2021, over 330 million cases were detected and 5.5 million deaths were recorded due to the pandemic (coronavirus.jhu.edu/map.html). In the United States, COVID-19 was the third leading cause of death in 2020 according to the National Center for Health Statistics [1]. Despite their socio-economic repercussions [2,3], non-pharmaceutical interventions such as social distancing, travel restrictions, and national lockdowns have proven to be effective at slowing the spread of the coronavirus [4–6]. As the pandemic evolved, pharmaceutical interventions, such as vaccinations and antiviral treatments, became increasingly important to manage the pandemic [7,8].

Less than a year into the pandemic, we witnessed the swift development of COVID-19 vaccines, expedited by new mRNA technology [9]. Both Pfizer-BioNTech [10] and Moderna [11] vaccines, among others, obtained emergency authorizations in the United States and Europe by the end of 2020, and governments began to distribute them to the public immediately. Mounting evidence shows that vaccines effectively prevent infections and severe hospitalizations, despite the emergence of new viral strains of the original SARS-CoV-2 virus [12,13]. It was estimated that the United States vaccination program averted up to 140,000 deaths by May 2021 [14] and over 10 million hospitalizations by November 2021 [15].

The widespread adoption of vaccines is extremely important to reduce the impact of the highly contagious virus [16]. However, as of December 2021 when supplies were no longer limited, only 62% of U.S. citizens had received two doses of COVID-19 vaccines [17]. Unvaccinated or partially vaccinated individuals still face risks of infection and death that are much higher than those who completed their vaccination cycle [18]. The geographically uneven vaccination coverage of the population can also lead to localized outbreaks and hinder governmental efforts to mitigate the pandemic [19].

Worldwide, most people are in favor of vaccines and vaccination programs, but a proportion of individuals are hesitant about some or all vaccines. Vaccine hesitancy describes a spectrum of attitudes, ranging from people with small concerns to those who completely refuse all vaccines. Previous literature links vaccine hesitancy to several factors that include the political, cultural, and social background of individuals, as well as their personal experience, education, and information environment [20]. Ever since public discourse moved online, concerns have been raised about the spread of false claims regarding vaccines on social media, which may erode public trust in science and promote vaccine hesitancy or refusal [21–24].

After the outbreak of the COVID-19 pandemic, a massive amount of health-related misinformation — the so-called "infodemic" [25] — was observed on multiple social media platforms [26–29], undermining public-health policies to contain the disease. Online misinformation included false claims and conspiracy theories about COVID-19 vaccines, hindering the effectiveness of vaccination campaigns [30,31].

A few recent studies reveal a positive association between exposure to misinformation and vaccine hesitancy at the individual level [32] as well as a negative association between the prevalence of online vaccine misinformation and vaccine uptake rates at the population level [33]. Motivated by these findings, our work investigates the spread of COVID-19 vaccine misinformation by analyzing almost 300 million English-language tweets shared during 2021, when vaccination programs were launched in most countries around the world.

There are a number of studies related to the present work. Yang et al. [29] carried out a comparative analysis of English-language COVID-19-related misinformation spreading on Twitter and Facebook during 2020. They compared the prevalence of low-credibility sources on the two platforms, highlighting how verified pages and accounts earned a considerable amount of reshares when posting content originating from unreliable websites. Muric et al. [34] released a public dataset of Twitter accounts and messages, collected at the end of 2020, which specifically focuses on anti-vaccine narratives. Preliminary analyses show that the online vaccine-hesitancy discourse was fueled by conservative-leaning individuals who shared a large amount of vaccine-related content from questionable sources. Sharma et al. [35] focused on identifying coordinated efforts to promote anti-vaccine narratives on Twitter during the first four months of the U.S. vaccination program. They also carried out a content-based analysis of the main misinformation narratives, finding that side effects were often mentioned along with COVID-19 conspiracy theories.

Our work makes two key contributions to existing research. First, we studied the prevalence of COVID-19 vaccine misinformation originating from low-credibility websites and YouTube videos and compared it to information published on mainstream news websites. As described above, previous studies either analyze the spread of misinformation about COVID-19 in general (during 2020) or focus specifically on anti-vaccination messages and narratives. They also analyze a limited time window, whereas our data captures 12 months into the roll-out of COVID-19 vaccination programs. Second, we uncovered the role and the contribution of important groups of vaccine misinformation spreaders, namely verified and automated accounts, whereas previous work either focuses on detecting users with a strong anti-vaccine stance or inauthentic coordinated behavior.

Considering these contributions, we address two research questions. The first is: **RQ1: What were the patterns of prevalence and contagion of COVID-19 vaccine misinformation on Twitter in 2021?** Leveraging a dataset of millions of tweets, we identified misinformation at the domain level based on a list of low-credibility sources (website domains) compiled by professional fact-checkers and journalists—an approach that is widely adopted in the literature to study unreliable information at scale [36–40]. Additionally, we considered a set of mainstream and public health sources as a baseline for reliable information. We then compared the volume of vaccine misinformation against reliable news, identified temporal trends, and investigated the most shared sources. We also explored the prevalence of misinformation that originated on YouTube and was shared on Twitter [29,41,42].

Analogously to the role of virus superspreaders in pandemic outbreaks [43], recent studies suggest that certain actors play an outsize role in disseminating misleading content [29,38,41]. For example, just 10 accounts were responsible for originating over 34% of low-credibility content shared on Twitter during an eight-month period in 2020 [44]. To examine how vaccine misinformation was posted and amplified by various actors on social media, our work addresses a second research question: **RQ2: Who were the main spreaders of vaccine misinformation?** Specifically, we analyzed two types of accounts. First, we investigated the presence and characteristics of users who generated the most reshares of misinformation [44,45], with a specific focus on the role of "verified" accounts. Twitter deems these accounts "authentic, notable, and active" (see help.twitter.com/en/managing-your-account/ about-twitter-verified-accounts). Second, we investigated the presence and role of social bots, i.e., social media accounts controlled in part by algorithms. Previous studies showed that bots actively amplified low-credibility information in various contexts [37,46,47].

**Table 1.** Sample keywords employed to collect tweets about vaccines.

| | | | |
|---|---|---|---|
| covid19vaccine | covidvaccine | coronavirusvaccine | vaccination |
| covid19 pfizer | pfizercovidvaccine | oxfordvaccine | getvaccinated covid19 |
| moderna | vaccine covid19 pfizer | mrna vaccinate covax | coronavirus |
| moderna | vax | | |

Our findings deepen our understanding of the ongoing pandemic and generate actionable knowledge for future health crises.

## Materials and methods

### Twitter data collection

On January 4th, 2021, we started a real-time collection of tweets about COVID-19 vaccines using the Twitter application program interface (API). The tweets were collected by matching relevant keywords through the *POST statuses/filter v1.1* API endpoint (developer.twitter.com/en/docs/twitter-api/v1/tweets/ filter-realtime/overview). This effort is part of our CoVaxxy project, which provides a public dashboard (osome.iu.edu/tools/covaxxy) to visualize the relationship between online (mis)information and COVID-19 vaccine adoption in the United States [48].

To capture the online public discourse around COVID-19 vaccines in English, we defined as complete a set as possible of English-language keywords related to the topic. Starting with *covid* and *vaccine* as our initial seeds, we employed a snowball sampling technique to identify co-occurring relevant keywords in December 2020 [48,49]. The resulting list contained almost 80 keywords. We show a few examples in Table 1; the full list can be accessed through the online repository associated with this project [50]. To validate the data collection procedure, we examined the coverage obtained by adding keywords one at a time, starting with the most common ones. Over 90% of the tweets contain at least one of the three most common keywords:
"vaccine," "vaccination," or "vaccinate." This indicates that the collected tweets are very relevant to the topic of vaccines.

In this paper, we analyzed the data collected in the period from January 4th to December 31st, 2021. This comprises 294,081,599 tweets shared by 19,581,249 unique users, containing 8,160,838 unique links (URLs) and 1,287,703 unique hashtags.

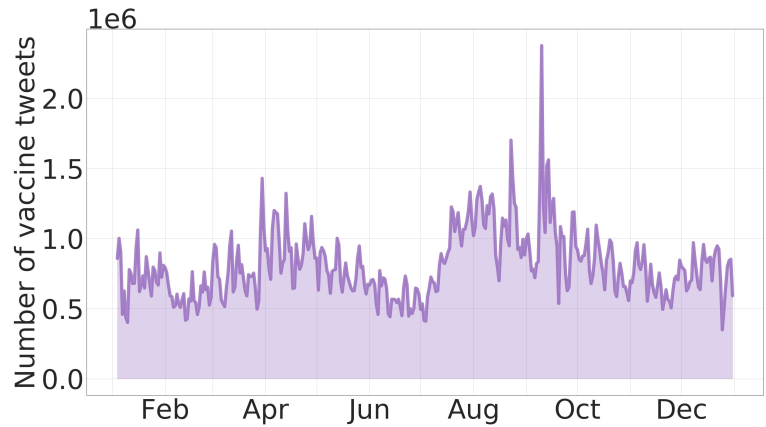

**Fig 1.** Time series of the daily number of vaccine-related tweets shared between January 4th and December 31st, 2021. The median daily number of tweets is 720,575.

Figure 1 shows the daily volume of vaccine tweets collected.

To comply with Twitter's terms of service, we are only able to share the tweet IDs with the public, accessible through a public repository [50]. One can "re-hydrate" the dataset by querying the Twitter API or using tools like *Hydrator* (github.com/DocNow/hydrator) or *twarc* (twarc-project.readthedocs.io/en/latest).

## Identifying online misinformation

We identified misinformation in our dataset using two approaches. Following a common method in the literature [36–40], the first approach identified tweets sharing links to low-credibility websites that were labeled by journalists, fact-checkers, and media experts for repeatedly sharing false news, hoaxes, conspiracy theories, unsubstantiated claims, hyperpartisan propaganda, click-bait, and so on. Specifically, we employed the Iffy+ Misinfo/Disinfo list of low-credibility sources (available at iffy.news/iffy-plus and accessed in March 2022). This list is mainly based on information provided by the Media Bias/Fact Check (MBFC) website (mediabiasfactcheck.com), an independent organization that reviews and rates the reliability of news sources. Political leaning was not considered for determining inclusion in the Iffy+ list. Instead, the list includes sites labeled by MBFC as having a "Very Low" or "Low" factual-reporting level and those classified as "Questionable" or "Conspiracy-Pseudoscience" sources. The 674 low-credibility sources in the Iffy+ list also include fake-news websites flagged by BuzzFeed, FactCheck.org, PolitiFact, and Wikipedia.

To expand our list of low-credibility sources, we also employed news reliability scores provided by NewsGuard [51], a journalistic organization that routinely assessed the reliability of news websites based on multiple criteria. NewsGuard assigns news outlets a trust score in the range [0,100]. While it considers outlets with scores below 60 as "unreliable," we adopted a stricter definition and only considered outlets with a score less than or equal to 30 as low-credibility. This yielded a list of 1,181 websites, which we cannot disclose to the public since the NewsGuard data is proprietary. By combining the Iffy+ list and the NewsGuard list, we obtained a total number of 1,718 low-credibility sources.

We tested the reliability of this domain-based approach to identify misinformation through a qualitative approach similar to previous studies [37,52]. We randomly chose 50 low-credibility links in our dataset and manually coded them as either "factual,"

**Table 2.** List of URL shortening services considered in our analysis.

| | | | |
|---|---|---|---|
| bit.ly | dlvr.it | liicr.nl | tinyurl.com |
| goo.gl | ift.tt | ow.ly | fxn.ws |
| buff.ly | back.ly | amzn.to | nyti.ms |
| nyp.st | dailysign.al | j.mp | wapo.st |
| reut.rs | drudge.tw | shar.es | sumo.ly |
| rebrand.ly | covfefe.bz | trib.al | yhoo.it |
| t.co | shr.lc | po.st | dld.bz |
| bitly.com | crfrm.us | flip.it | mf.tt |
| wp.me | voat.co | zurl.co | fw.to |
| mol.im | read.bi | disq.us | tmsnrt.rs |
| usat.ly | aje.io | sc.mp | gop.cm |
| crwd.fr | zpr.io | scq.io | trib.in |
| owl.li | | | |

"misinformation," or "unverified." Two authors independently visited the actual web page of each link and researched its content to determine if it was accurate. A link was coded as "factual" if all claims within the article were corroborated by other sources. The "unverified" label was utilized for links that could no longer be accessed (e.g., because the web page no longer exists). All other links were coded as "misinformation." In the event of coding disagreements, authors shared and discussed what they learned during their independent research to reach an agreement on a single label. At the end of this procedure, seven links were coded as "factual," 38 as "misinformation," four as "unverified," and a single article was excluded as it appeared to be a personal blog post. We also note that of the seven articles labeled as "factual," six were from state propaganda outlets with a selection bias (e.g., sputniknews.com or rt.com).

As a second approach, we analyzed links to YouTube videos shared on Twitter that might contain misinformation. We extracted unique video identifiers from links shared in the collected tweets and queried the YouTube API for the video status using the *Videos:list* endpoint. In light of recent YouTube efforts to remove anti-vaccine videos according to their COVID-19 policy [53] and their updated policy [54], we considered videos to be suspicious if they were not publicly accessible. Previous research shows that inaccessible videos contain a high proportion of anti-vaccine content, such as the "Plandemic" conspiracy documentary [29]. The efficacy of this approach to identifying videos that contain anti-vaccine content is further supported in research that analyzed available videos shared by users that had also shared an inaccessible video [55]. The authors found that the majority of available videos tweeted by these users promulgated an anti-vaccine or anti-mandate stance. As some estimates suggest that it takes an average of 41 days for YouTube to remove videos that violate their terms [42], we checked the statuses of videos in March 2022, at least 2 months after the last video was posted on Twitter.

## Sources of reliable information

We curated a list of reliable, mainstream sources of vaccine-related news as our baseline to interpret the prevalence of misinformation and characterized its spreading patterns [29]. In particular, we considered websites with a NewsGuard trust score higher than 80, resulting in a list of 2,765 sources. We also included the websites of two authoritative sources of COVID-19-related information, namely the U.S. Centers for Disease Control and Prevention (cdc.gov, CDC) and the World Health Organization (who.int, WHO). In the rest of the paper, we use "low-credibility" and "mainstream" to refer to the two sets of sources.

### Link extraction

Identifying low- and high-credibility links and YouTube links requires extracting the top-level domains from the URLs embedded in tweets and matching them against our lists of web domains. Shortened links occurred frequently in our dataset, therefore we identified the most prevalent link-shortening services (the list can be found in Table 2) and obtained the original links through HTTP requests.

### Bot detection

To measure the level of bot activity for different types of information, we employed BotometerLite (accessible at rapidapi.com/OSoMe/api/botometer-pro), a publicly-available tool that can efficiently identify likely automated accounts on Twitter [56]. For each Twitter account, BotometerLite generates a bot score in the range [0,1] where a higher score indicates that the account is more likely to be automated. BotometerLite evaluates an account by inspecting the profile information that is embedded in each tweet. This enabled us to perform bot analysis at the level of tweets in our dataset.

### Ethical considerations

This research is based on observations of public data with minimal risks to human subjects. It was deemed exempt from review by the Indiana University IRB (protocol 1102004860). Data collection and analysis are performed in compliance with the terms of service of Twitter.

## Results

### Prevalence and contagion of online misinformation

To address **RQ1**, we compared the prevalence of tweets that linked to domains in our lists of low-credibility and mainstream sources over time. We carried out a similar analysis for suspicious YouTube videos. As shown in panels **A** and **B** of Fig. 2, we observed a significant increasing trend in the daily prevalence of low-credibility information over time and a significant opposite trend for mainstream news. This is further confirmed in panel **C**, which shows the daily ratio between the volumes of tweets linking to low-credibility and mainstream news. A significant increasing trend was observed, suggesting that the public discussion about vaccines on Twitter shifted over time from referencing trustworthy sources in favor of low-credibility sources. The peak in July corresponds to a time when the prevalence of mainstream news was particularly low (panel **B**). During this period we also observed a burst of reshares for content originating from Children's Health Defense, the most prominent source of vaccine misinformation (further discussed below).

    During the entire period of analysis, we found that misinformation is generally less prevalent than mainstream news, as shown in panel **A** of Fig. 3. However, we observed that low-credibility content tended to spread more through retweets compared to mainstream content, as shown in panel **B** of Fig. 3. This indicated that while low-

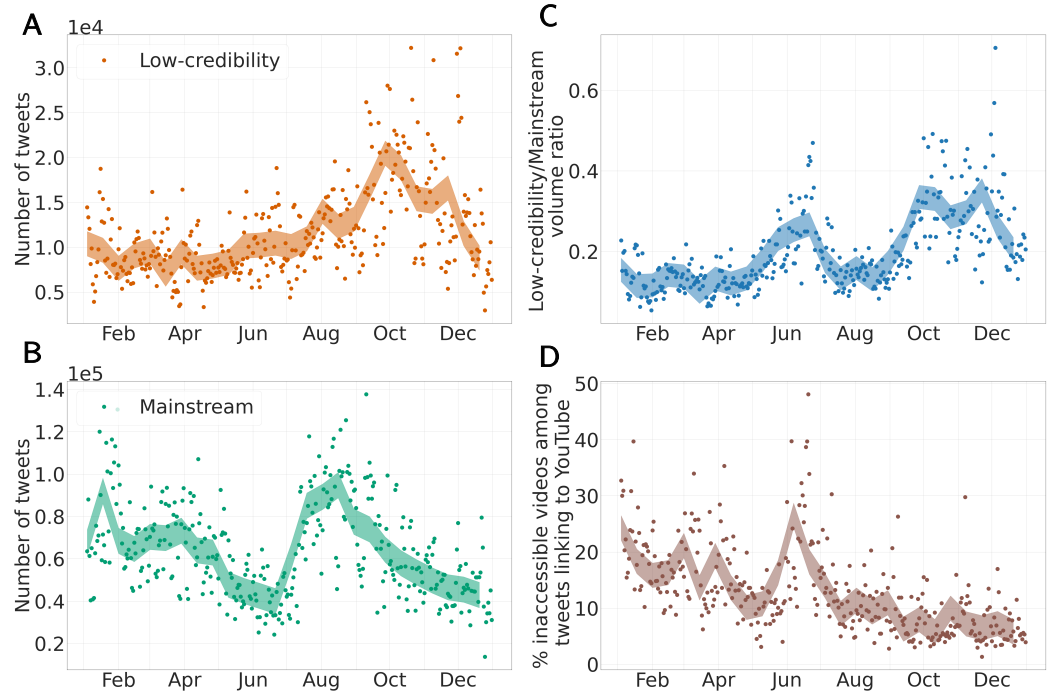

**Fig 2.** Timelines of prevalence of vaccine information on Twitter. We employ non-parametric Mann-Kendall tests for trends. Colored bands correspond to a 14-day rolling average with 95% C.I. **(A)** Daily number of vaccine tweets sharing links to news articles from low-credibility sources. There is a significant increasing trend ($P < .001$). **(B)** Daily number of vaccine tweets sharing links to news articles from mainstream sources. There is a significant decreasing trend ($P < .001$). **(C)** Ratio between the volumes of tweets sharing links to low-credibility and mainstream sources. There is a significant increasing trend ($P < .001$). **(D)** Daily percentage of tweets sharing links to inaccessible YouTube videos, out of all tweets sharing links to YouTube. There is a significant decreasing trend ($P < .001$).

credibility vaccine content was less prevalent overall, it had a greater potential for contagion through the social network, suggesting that it might have only spread through a subsection of the population.

We further report that the fraction of vaccine-related tweets linking to YouTube videos was very small (daily median: 0.52%). However, a non-negligible proportion of these posts (daily median: 10.95%) shared links to inaccessible videos, with a larger prevalence in the first half of 2021 (a peak of 45% is observed in July), and a significant decreasing trend towards the end of the year (see panel **D** of Fig. 2).

## Most popular misinformation sources

Looking at different sources of news about vaccines, panel **A** in Fig. 4 shows the 20 most shared websites. We note three unreliable sources in this ranking, namely childrenshealthdefense.org, thegatewaypundit.com, and zerohedge.com. The most popular low-credibility source was the website of the Children's Health Defense (CHD) organization, an anti-vaccine group led by Robert F. Kennedy Jr. that became very popular during the pandemic as an alternative and natural medicine site [45,57]. This source was banned from Facebook and Instagram for repeatedly violating their guidelines against spreading medical misinformation in August 2022 [58]. With around 0.30% of all vaccine tweets, its prevalence was comparable to that of reputable sources such as washingtonpost.com and reuters.com, and roughly twice the prevalence of CDC links (0.16%). As shown in panel **B**, CHD was much more widely shared than other low-credibility sources, most of which had less than 0.05% of all shared tweets. CHD accounted for approximately 18% of all tweets linking to low-credibility sources, whereas the aggregated 20 most shared sources generated around 61% of all such tweets. Nevertheless, the total fraction of tweets sharing low-credibility news about vaccines accounted for only 1.5% compared to approximately 7.8% of tweets that linked to mainstream sources (see panel **C** of Fig.4).

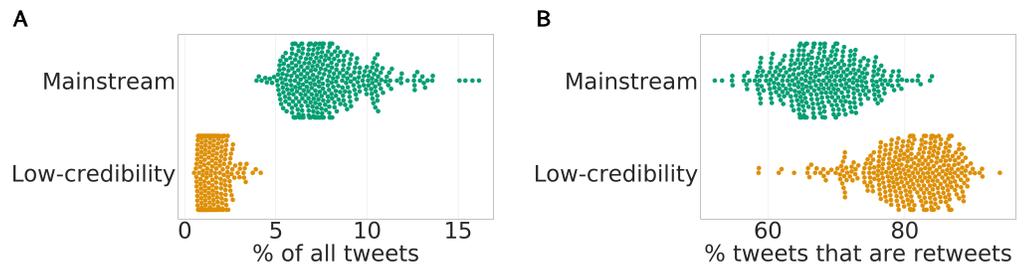

**Fig 3.** Comparisons between prevalence of tweets linking to mainstream and low-credibility sources. **(A)** Daily percentage of vaccine tweets and retweets that share links to low-credibility news sources (median: 1.31%) and mainstream news sources (median: 7.53%). The distributions are statistically different according to a two-sided Mann-Whitney test ($P < .001$). **(B)** Distributions of the proportion of tweets linking to low-credibility sources (median: 89.19%) and mainstream sources (median: 67.96%) that are retweets. The distributions are statistically different according to a two-sided Mann-Whitney test ($P < .001$).

## Superspreaders of misinformation

Recent work reveals that accounts who disseminated a disproportionate amount of low-credibility content—so-called "superspreaders"—played a central role in the digital misinformation crisis [29,38,41,44,45]. These contributions also show that "verified" accounts often act as superspreaders of unreliable information, therefore we further investigated the role of such accounts to address **RQ2**.

Figure 5 shows that over time, verified accounts represented around 15% of those that posted vaccine content, but were consistently responsible for about 43% of that content. When we focus on low-credibility content, verified accounts represented an even smaller proportion of accounts, less than 6%. Still, they were responsible for approximately 34% of retweets. These findings highlight a stunning concentration of impact and responsibility for the spread of vaccine misinformation among a small group of verified accounts. While there were substantially fewer verified accounts sharing low-credibility vaccine content (828) compared to those sharing vaccine content in general (98,612), Figure 6 shows that verified accounts tended to receive more retweets when posting low-credibility content than general vaccine content.

January 17, 2023                                                                                                          9/20

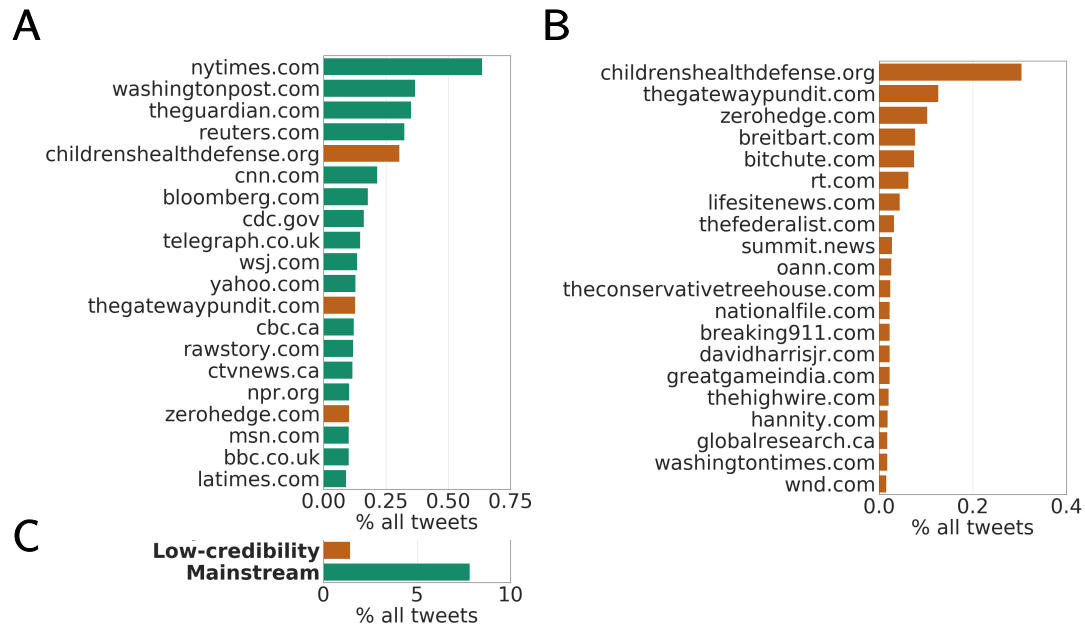

**Fig 4.** Top sources of vaccine content. **(A)** The top 20 news sources ranked by percentage of vaccine tweets. **(B)** The top 20 low-credibility news sources ranked by percentage of vaccine tweets. **(C)** Percentages of all vaccine tweets linking to low-credibility and mainstream news sources.

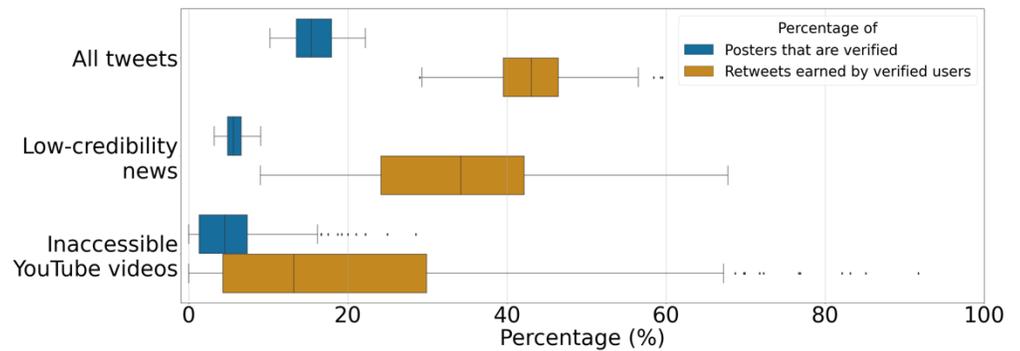

**Fig 5.** Comparisons between percentages of original posters who are verified accounts and of retweets earned by verified accounts, for different categories of vaccine content. Each data point is a daily proportion. The median daily proportions of verified accounts among posters of vaccine content, low-credibility news, and inaccessible YouTube videos are 15.4%, 5.6%, and 4.5%, respectively. The median daily proportions of retweets earned by verified posters of vaccine content, low-credibility news, and inaccessible YouTube videos are 43.1%, 34.2%, and 13.2%, respectively. All distributions are statistically different from each other according to two-sided Mann-Whitney tests ($P < 0.001$).

In Fig. 7 we ranked the top 25 accounts by the number of retweets to their posts linking to low-credibility sources. 11 of these misinformation superspreaders were accounts that have been verified by Twitter, some of which are associated with untrustworthy news sources (e.g., *@zerohedge*, *@BreitbartNews*, and *@OANN*). The top superspreader, Robert Kennedy Jr. (*@RobertKennedyJr*), earned approximately 3.45 times the number of retweets of the second most-retweeted account (*@zerohedge*). Mr. Kennedy was identified as one of the pandemic's "disinformation dozen" [41,45]. His influence fueled the high prevalence of links to childrenhealthdefense.org within our dataset (as previously shown in Fig. 4). His verified account had approximately 3.8 times more followers than the unverified *@ChildrensHD* account (416.2k versus 109.8k, respectively as of April 24th, 2022). Retweets of Mr. Kennedy's tweets singularly accounted for 13.4% of all retweets of low-credibility vaccine content. A robustness check removing this account from the data yielded consistent results for all analyses reported in this section.

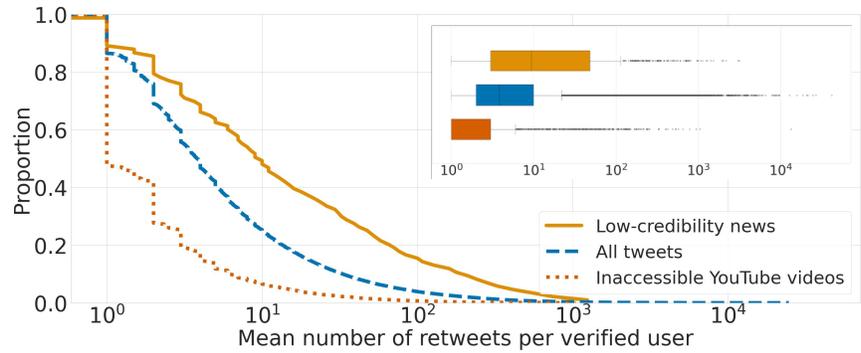

**Fig 6.** Distributions of the mean numbers of retweets earned by verified accounts when sharing vaccine content (median 3.82), low-credibility news (median 9.43), and links to inaccessible YouTube videos (median 1). We display the complementary cumulative distributions in the main plot because the distributions are broad. In fact, the box plots (inset) have many outliers. All distributions are significantly different from each others according to two-sided Mann-Whitney tests ($P < 0.001$).

We also investigated the role of verified users in sharing suspicious videos from YouTube. As shown in Figs. 5 and 6, we found that verified accounts do not play as central a role in spreading this content in contrast to content from low-credibility domains.

## Role of social bots

To address **RQ2**, we also inspected the role of likely automated accounts in spreading COVID-19 vaccine misinformation. As mentioned in the Methods section, we employed BotometerLite [56] to calculate a bot score for all the accounts posting a tweet in our dataset. We did not observe notable temporal trends in the activity of likely bots over time, so we show in Fig. 8 the distributions of daily average bot scores for tweets sharing vaccine content, links to low-credibility sources, and inaccessible YouTube videos.

We observed that tweets sharing links to low-credibility sources had significantly higher bot-activity levels than vaccine tweets overall. As for tweets sharing inaccessible YouTube videos, their daily average bot scores were even higher than those linking to low-credibility sources.

This analysis was carried out at the tweet level, meaning that if a bot-like account tweeted more times, it made a larger contribution. We observed similar results when performing the analysis at the account level, by considering the contribution of each account once.

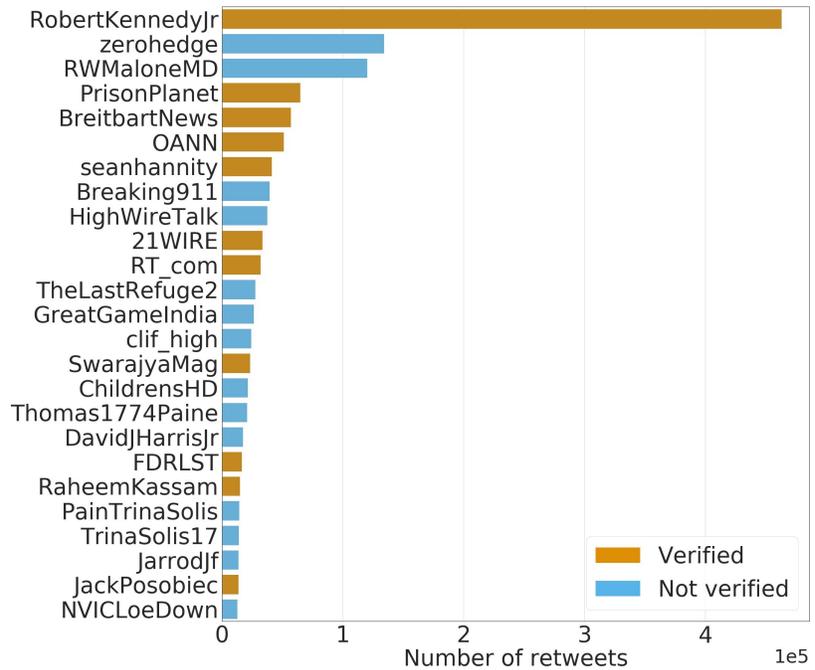

**Fig 7.** Top 25 accounts ranked by the number of retweets earned when sharing links to low-credibility news websites. Colors indicate whether accounts are verified (orange) or not (blue).

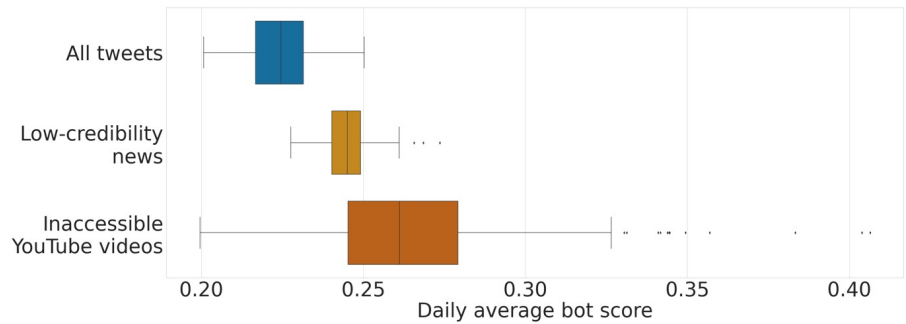

**Fig 8.** Comparison between the daily average bot score of tweets sharing different categories of vaccine content. The median daily average bot scores of accounts sharing vaccine content, low-credibility news, and inaccessible YouTube videos are 0.22, 0.25 and 0.26, respectively. All distributions are significantly different from each other according to two-sided Mann-Whitney tests ($P < 0.001$).

## Discussion

We investigated COVID-19 vaccine misinformation spreading on Twitter during 2021 following the roll-out of vaccination programs around the world. Leveraging a source-based labeling approach, we identified millions of tweets sharing links to low-credibility and mainstream news websites. While low-credibility information was generally less prevalent than mainstream content over the year, we observed an increasing trend in the reshares of unreliable news during the year, and an opposite, decreasing trend for reliable information. Our data mostly captures English-language conversations, which could originate from different countries. However, our aggregate analysis does not disentangle the infodemic trends and peaks associated with different countries, as observed in prior work [28].

Focusing on specific news sources, we noticed three low-credibility websites with volumes of reshares comparable to reliable sources. Alarmingly, the most prominent source of vaccine misinformation, Children's Health Defense, earned more than twice the number of tweets linking to the Centers for Disease Control and Prevention. Looking at users who earned the most retweets when sharing low-credibility news about vaccines, we observed the presence of many verified accounts. In particular, the verified user who earned the most retweets was Robert Kennedy Jr., the founder of Children's Health Defense.

Given the increase in misinformation over time and the outsized role of a small group of verified users, we hypothesize that financial incentives may play an important role [57,59]. Low-credibility websites monetize visitors through donations, advertising, and merchandise. Our finding that vaccine misinformation tended to spread more through retweets compared to mainstream news suggests that misinformation content lends itself to such exploitation. Amplification by automated accounts may also have played a role in increasing levels of misinformation, as we found these accounts to be significantly more active at sharing low-credibility news and inaccessible YouTube videos compared to vaccine-related content overall. However, we did not find a trend of increased levels of automated sharing over time.

There are a number of limitations to our study. The source-based approach to identify low-credibility information at scale is not perfect. Credible sources may occasionally report inaccuracies and low-credibility sources often publish a mixture of reliable and unreliable information. Our analysis based on a sample of articles suggests that approximately 76% of articles from low-credibility sources do contain false or misleading content.

While we cannot publicly disclose Newsguard ratings, they are available to researchers upon agreement, and this should ensure reproducibility. We elected to include Newsguard data because it is more comprehensive, up-to-date, and its methodology is better documented compared to other ratings like those from Iffy+ Misinfo/Disinfo. Nevertheless, it would be possible to repeat our analysis using only free ratings from Iffy+ Misinfo/Disinfo since, as the literature suggests, the ratings are highly correlated [60,61]. In fact, we observe a high overlap between our lists of top sources — for example, 17 of the top-20 sources in Figure 4B are also present in the Iffy+ Misinfo/Disinfo list. More importantly, over 86% of the total number of low-credibility tweets identified with the merged list originate from websites contained in the Iffy+ Misinfo/Disinfo list alone. This suggests that the results are robust with respect to the ratings source.

Similar limitations exist with respect to labeling inaccessible YouTube videos as low-credibility. For example, some of these videos may be inaccessible due to restricted access or copyright violations. An uploader's choice to restrict access to a video may serve as a way to circumvent content moderation policies or could be unrelated to anti-vaccination efforts. However, in the context of the vaccination discussion on Twitter, examinations of videos and their Twitter posters suggest that most inaccessible videos are likely anti-vaccination in their orientation [55]. In addition, not all accessible videos contain accurate information about vaccines. YouTube may fail to identify content that should be removed according to its own policies. As such, analyses of inaccessible videos should be treated more like lower-bound estimates.

Another limitation is that the Botometer algorithm we employ to detect automated accounts is not perfect and may not accurately classify social bots [56]. We investigated whether bot-like behavior, as identified by Botometer, is associated with suspicious activity on Twitter. We used Twitter's Compliance API (developer.twitter.com/en/docs/twitter-api/compliance/batch-compliance/introduction) to check all accounts for suspension from the platform as of November 2022. We observed a significant positive correlation between the BotometerLite score, binned into 40 equal intervals, and the proportion of accounts suspended (Pearson's $R$ = 0.93, $P$ < 0.001). This suggests that the classifier reliably reveals behaviors that eventually lead to suspension on the platform. Perhaps most importantly, Twitter users may not be very representative of the real-world population across a range of demographic groups [62], although information circulating around Twitter can have a great influence over the news media agenda [63]. Further studies should consider multiple social media platforms simultaneously, especially those with upward adoption trends [64].

Despite these limitations, our findings help map the landscape of online vaccine misinformation and design intervention strategies to curb its spread. The presence of misinformation around COVID-19 vaccines on Twitter shows that there was an audience for this type of content, which might reflect a deeper distrust of medicine, health professionals, and science [65]. In a context of widespread uncertainty such as the COVID-19 pandemic, trust is critical for overcoming vaccine hesitancy, and recent research shows how online misinformation fueled vaccine hesitancy and refusal sentiment [24,33].

Our findings reveal the presence of a small number of main producers and repeat spreaders of low-credibility content. Given that these superspreaders played key roles in disseminating vaccine misinformation, a straightforward strategy could be to deplatform them [66,67], as shown by recent studies in different contexts [67–69] — and as has been done by major platforms in notable cases such as Alex Jones [70] and Donald Trump (blog.twitter.com/en_us/topics/company/2020/suspension).

While social media platforms have legal rights to regulate online conversations, the decisions to deplatform public figures should be made with caution. In fact, past interventions have sparked a vivid debate around free speech and caused many users to migrate to alternative platforms [67,69]. It is also unclear whether reducing the supply of false information and increasing the supply of accurate information can "cure" the problem of vaccine hesitancy [31]. An alternative path of action could be to reduce the financial incentives of those who profit from the spread of misinformation. Our results also show that vaccine misinformation is more viral than other kinds of information. Other effective approaches to reduce its spread include lowering the visibility of certain content ("down-ranking") or not showing that content to users ("shadow banning"), as well as adding warning labels to content that is potentially harmful or inaccurate [71,72]. Platforms should partner with policymakers and researchers in evaluating the impacts of such different interventions [73].

There are several interesting research questions that are outside the scope of the present work, but that could be addressed by future research. For instance, further investigations could address the impact of Twitter's removal of users due to the January 6$^{th}$ riots on the spread of misinformation in the following months. Other studies could investigate how the Children's Health Defense organization shifted its anti-vaccination narratives from children to a broader COVID-19 vaccination campaign and remained the most popular source for the anti-vaccination movement. Future work could also analyze *exposure* to low-credibility information, which is more difficult to measure compared to the *sharing patterns* quantified in this paper. This would allow answering the question of whether the spread of low-credibility information was confined to a limited group of people or reached a wide audience. Finally, it is still unclear how governmental and societal changes might have affected conversations around vaccines during the COVID-19 pandemic compared to the (anti-)vaccination debate in previous years.

All in all, we believe our work provides actionable insights for addressing the online spread of vaccine misinformation. Such insights can be beneficial during the ongoing pandemic and future health crises.

## Acknowledgements

This work is supported in part by the Italian Ministry of Education (PRIN project HOPE), the EU Horizon 2020 (grant 101016233), the National Science Foundation (grant ACI-1548562), Knight Foundation, and Craig Newmark Philanthropies.